\documentclass[
reprint,longbibliography,
groupedaddress,
 amsmath,amssymb,
prl,
]{revtex4-2}

\newcommand\beq{\begin{equation}}
\newcommand\eeq{\end{equation}}

\usepackage{graphicx}
\usepackage{amsmath}
\usepackage{amsfonts}
\usepackage{color}
\usepackage[normalem]{ulem}
\usepackage[colorlinks, linkcolor= blue, citecolor = blue, urlcolor=blue]{hyperref}

\begin{document}


\title{Exploring Aperiodic Order in Photonic Time Crystals}

\author{Marino Coppolaro}
\affiliation{Fields \& Waves Lab, Department of Engineering, University of Sannio, I-82100 Benevento, Italy}
\author{Massimo Moccia}
\affiliation{Fields \& Waves Lab, Department of Engineering, University of Sannio, I-82100 Benevento, Italy}
\author{Giuseppe Castaldi}
\affiliation{Fields \& Waves Lab, Department of Engineering, University of Sannio, I-82100 Benevento, Italy}
\author{Vincenzo Galdi}
\email{vgaldi@unisannio.it}
\affiliation{Fields \& Waves Lab, Department of Engineering, University of Sannio, I-82100 Benevento, Italy}

\date{\today}


\begin{abstract}
We present a theoretical framework for analyzing aperiodically ordered photonic time quasicrystals (PTQCs), which are the temporal analogs of spatial photonic quasicrystals. Using a general two-symbol substitutional sequence to model temporal modulations, we extend the trace and anti-trace map formalism  used for spatial photonic quasicrystals to the temporal domain. Focusing on the Thue-Morse sequence as a representative example, we examine the band structure and wave-transport properties, discussing their physical origins and highlighting both similarities and key differences with conventional periodic photonic time crystals. Furthermore, we investigate the peculiar features of PTQCs, such as multiscale spectral response and localization effects. Our findings provide valuable insights into the complex interplay between aperiodic order and wave dynamics in time-varying media, highlighting its potential to enable the development of advanced photonic devices.
\end{abstract}

\maketitle

 %
 \begin{figure}[t]
 	\centering
 	\includegraphics[width=\linewidth]{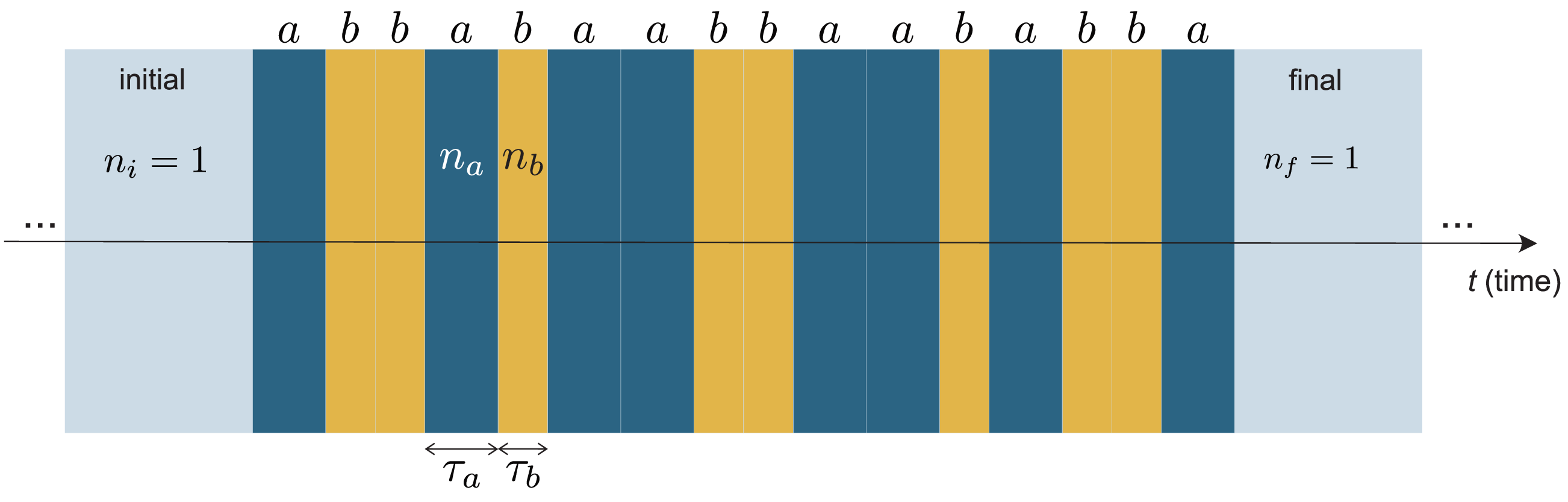}
 	\caption{Problem schematic (details in the text). The specific arrangement corresponds to a Thue-Morse sequence at iteration order $n=4$.}
 	\label{Figure1}
 \end{figure}
 
 The resurgence of interest  in time-varying electromagnetic media \cite{Galiffi:2022po} builds upon foundational studies from the 1950s \cite{Morgenthaler:1958vm, Oliner:1961wp, Felsen:1970wp, Fante:1971to}, which first investigated the impact of temporally modulated material properties. While these early works were predominantly theoretical, recent advances in ultrafast technologies and metamaterials engineering have transformed these concepts into practical possibilities. Often referred to as 4-D metamaterials \cite{Engheta:2023fd}, these dynamic systems enable unique functionalities, including nonreciprocal wave responses, joint manipulation of spatial and spectral properties, and the ability to transcend fundamental limitations of linear, time-invariant media \cite{Hayran:2023ut}.
 
Among the most intriguing developments in this field are ``photonic time crystals'' (PTCs) \cite{Romero:2016tp}, which  have emerged as a prominent area of study.  These structures can be regarded as the temporal analogs of photonic crystals, with their constitutive parameters undergoing temporal modulations on time scales comparable to the period of electromagnetic waves. Similar to their spatial counterparts, PTCs exhibit strong dispersion, but their band structures are defined by {\em momentum} (wavevector) bandgaps rather than {\em frequency} bandgaps. Remarkably, due to the non-conservation of energy in these systems, momentum bandgaps are linked to exponential signal amplification, as energy is extracted from the temporal modulation. These unique properties open up exciting possibilities in both fundamental research and practical applications, including topological photonics \cite{Lustig:2018ta}, thresholdless lasers \cite{Lyubarov:2022ae, Dikopoltsev:2022le}, free-electron radiation \cite{Dikopoltsev:2022le}, and optical absorbers \cite{Hayran:2024bt}. For recent reviews and perspectives, including potential material platforms, applications and ongoing challenges, the reader is referred to \cite{Lustig:2023pt, Saha:2023pt, Boltasseva:2024pt, Asgari:2024ta}.
 
While most of the existing literature on PTCs has focused on {\em periodic} temporal modulations, a few notable studies have explored the impact of {\em random disorder} \cite{Sharabi:2021dp, Eswaran:2024al} and different forms of non-periodic modulations \cite{Nemirovsky:2023pt,Koufidis:2024es,Ni:2025tw,Araujo:2025op}. These works suggest that moving away from strict periodicity can offer significant benefits, revealing new phenomena such as Anderson-like localization and topological effects, and improving performance, such as enabling higher and broader pulselike amplification.
 
Interestingly, the above observations parallel the trajectory of research in spatial photonic crystals, where the concept of ``quasicrystals'' \cite{Shechtman:1984mp}, i.e., structures that exhibit {\em aperiodic} yet non-random order, has unveiled  novel optical properties and design opportunities, including enhanced light localization and advanced bandgap engineering \cite{Macia:2012ea, Vardeny:2013oo}. This suggests that extending similar principles to PTCs, through a systematic exploration of aperiodic temporal modulations, could lead to equally rich and transformative phenomena.

To this aim, we introduce a rigorous theoretical framework for modeling aperiodically ordered PTCs based on general two-symbol substitutional sequences, employing the trace and anti-trace map formalism \cite{Kohmoto:1983lp, Kolar:1990tm, Wang:2000ta}. This methodology allows for the exploration of temporal analogs to well-known spatial photonic quasicrystals, such as those derived from Thue-Morse or Fibonacci sequences. Similar to their spatial counterparts, these  ``photonic time quasicrystals'' (PTQCs) exhibit complex band structures and gap-edge states, which may display fractal scaling and anomalous localization.

As schematized in Fig. \ref{Figure1}, we consider a medium that is unbounded in space and, starting from a stationary state, ideally undergoes step-like temporal modulations of its constitutive properties. The modulation alternates between intervals of duration $\tau_a$ with refractive index $n_a$, and intervals of duration $\tau_b$ with refractive index $n_b$, with a rule described hereafter. Without loss of generality, we assume that the initial and final (stationary) states of the medium are identical and correspond to vacuum, i.e., $n_i = n_f = 1$. Furthermore, to isolate the effect of aperiodic temporal modulation, we neglect dispersion and losses, assuming operation far from material resonances.

The temporal modulation is ruled by a general two-symbol substitution rule:
\beq
a\rightarrow \alpha\left(a,b\right),\quad b\rightarrow \beta\left(a,b\right),
\label{eq:SR}
\eeq
with $\alpha$ and $\beta$ denoting strings of symbols ``$a$'' and ``$b$''. Starting from a suitable initialization, the sequence is generated by inflation, repeatedly applying the substitution rules (\ref{eq:SR}) until the desired length is reached.

We focus on studying the propagation of plane waves in the resulting PTQC, characterized by a spatial dependence of $\exp(i k x)$, 
where $k$ denotes the conserved wavenumber and $x$ represents the  propagation direction (chosen without loss of generality).
 Analogous to the spatial counterpart, this propagation can be analytically modeled, with each of the two types of modulation intervals described by a unimodular {\em transfer matrix}:
\beq
{\underline{\bf S}}_{\nu}(k) =
\begin{bmatrix}
	\cos \varphi_\nu & -\displaystyle{\frac{1}{n_\nu}} \sin \varphi_\nu \\
	n_\nu \sin \varphi_\nu & \cos \varphi_\nu
\end{bmatrix}, \quad \nu = a, b,
\label{eq:Mm}
\eeq
which relates the electric and magnetic displacement field components (chosen as $D_y$ and $B_z$, respectively, without loss of generality) at the two ends (see \cite{SM} for further details). In  (\ref{eq:Mm}), $\varphi_\nu = ck \tau_\nu/n_\nu$ represents the normalized travel time, and $c$ is the speed of light in vacuum.
Accordingly, at a given iteration order $n$ of the substitution rules (\ref{eq:SR}), the entire PTQC can be described by a transfer matrix obtained via the backward chain product:
\beq
{\underline{\bf M}}_n=\prod_{\ell=L_n}^{1} {\underline{\bf S}}_{\nu(\ell)}=
\begin{bmatrix}
	M^{(n)}_{11} & M^{(n)}_{12}\\
	M^{(n)}_{21} & M^{(n)}_{12}
\end{bmatrix},
\label{eq:MM}
\eeq
where $L_n$ represents the length of the $n$th-order sequence, and $\nu(\ell)$ denotes the type (``$a$'' or ``$b$'') of the $\ell$th modulation interval.
It is important to note that the definition of the transfer matrix is not unique. Remarkably, our specific formulation enables a compact and insightful characterization of the key properties of a PTQC in terms of the (real-valued) {\em trace} ($\mbox{Tr}$) and {\em anti-trace} ($\mbox{ATr}$) of the transfer matrix:
\begin{subequations}
	\begin{eqnarray}
		\chi_n&\equiv& \mbox{Tr}\left({\underline{\bf M}}_n\right)=M^{(n)}_{11}+M^{(n)}_{22},\\
		\upsilon_n&\equiv& \mbox{ATr}\left({\underline{\bf M}}_n\right)=M_{21}^{(n)}-M_{12}^{(n)}.
	\end{eqnarray}
\end{subequations}
Specifically, for a PTQC of {\em infinite} duration obtained via periodic replication of an $n$th-order ``supercell'', the dispersion relation connecting the angular eigenfrequency $\Omega$ and wavenumber can be written as (see \cite{SM} for  details):
\beq
\cos\left(\Omega \tau_n\right)=\frac{\chi_n}{2},
\label{eq:DE}
\eeq
where $\tau_n$ represents the duration of the $n$th-order supercell. From (\ref{eq:DE}), it is clear that the band structure is entirely determined by the trace $\chi_n$. Specifically, 
momentum bandgap (i.e., complex-valued $\Omega$) are characterized by the condition $\left|\chi_n\right|>2$.

On the other hand, for a PTQC of {\em finite} duration, the relevant observables are the reflection (backward-wave) and transmission (forward-wave) coefficients, denoted by ${\cal R}_n$ and ${\cal T}_n$, respectively. It can be shown (see \cite{SM} for details) that their magnitudes are entirely determined by the trace and anti-trace:
\begin{subequations}
	\begin{eqnarray}
		\left|{\cal T}_n\right|^2&=&\frac{\chi_n^2+\upsilon_n^2}{4},
		\label{eq:Tn}\\
		\left|{\cal R}_n\right|^2&=& \left|{\cal T}_n\right|^2-1.
		\label{eq:Rn}
	\end{eqnarray}
	\label{eq:TR}
\end{subequations}
Moreover, starting from (\ref{eq:Tn}), it can be shown (see \cite{SM} for details) that $\left|{\cal T}_n\right| \ge 1$.
This does not contradict the presence of nonzero reflections, as power is not generally conserved in our system, as implied by (\ref{eq:Rn}).
In the following, we focus on studying the transmission response, as the reflection response differs only by a constant pedestal.

The proposed approach is particularly powerful as it can accommodate {\em any} substitutional sequence, including the conventional periodic case (see \cite{SM} for details and examples). Trace and anti-trace maps can be derived, in principle, for {\em arbitrary} sequences, even those involving more than two symbols \cite{Kolar:1990tm,Wang:2000ta}. This approach presents two key advantages: first, it enables a computationally efficient and insightful analysis of PTQC properties, eliminating the need for the cumbersome matrix product in (\ref{eq:MM}); and second, it offers a clear and intuitive perspective on phenomena across multiple time scales, while effectively identifying and parameterizing the key design factors.  Furthermore, this framework facilitates leveraging established results from the study of spatial photonic quasicrystals, extending their relevance to the temporal domain.

As a representative example, we focus on the Thue-Morse sequence \cite{Queffelec:2010sd}, which is defined by the substitution rules:
\beq
\alpha\left(a,b\right)=ab,\quad \beta\left(a,b\right)=ba.
\eeq
Upon iteration, this generates the sequences: $ab$, $abba$, $abbabaab$, and so on, with length $L_n=2^n$. For instance, the PTQC geometry shown in Fig. \ref{Figure1} is based on the $n=4$ order of this sequence.
The Thue-Morse sequence captures several key aspects of aperiodic order, including a {\em singular-continuous} Fourier spectrum, which displays a fractal-like, continuous distribution of frequencies without sharp peaks or discrete states \cite{Queffelec:2010sd}.
 Besides being extensively studied in the context of spatial photonic quasicrystals \cite{Liu:1997po,Jiang:2005pg,Lei:2007pb,Grigoriev:2010bm,Savoia:2013on,Coppolaro:2018ao,Zhang:2021ft}, this geometry is particularly well-suited for direct comparison with the periodic scenario. It maintains the same proportions of type $``a"$ and $``b"$ constituents while differing in their arrangement, making it ideal for isolating and analyzing the effects of positional order.

For a Thue-Morse sequence, the trace and anti-trace maps can be written as \cite{Liu:1997po}:
\begin{subequations}
	\begin{eqnarray}
			\chi_{n}&=&\chi_{n-2}^2\left(\chi_{n-1}-2\right)+2,\\
		\upsilon_{n}&=&
		\upsilon_{n-2}\chi_{n-2}\chi_{n-1}\nonumber\\
		&+&\left[
		1+(-1)^n
		\right]
		\left(
		\upsilon_{n-1}-\upsilon_{n-2}\chi_{n-2}
		\right)
		,\quad n\ge 3.
	\end{eqnarray}
	\label{eq:ThMmaps}
\end{subequations}
These maps are initialized with the values for $n = 1, 2$, which can be easily computed analytically (see \cite{SM} for details), as the first two iterations are simply periodic sequences with unit cells of type ``$ab$" and ``$abba$''. 

By simple inspection of the maps in (\ref{eq:ThMmaps}), we can observe that $\chi = 2$ and $\upsilon = 0$ form a {\em fixed point}. From (\ref{eq:DE}) and (\ref{eq:TR}), it is evident that this condition corresponds to the band edges in the infinite case, and to perfect transmission (${\cal T} = 1$, ${\cal R} = 0$) in the finite case.

For our parametric studies, we set $n_a=1$ and $n_b=1.5$, and define a reference modulation period $T_0$ (and corresponding angular frequency $\Omega_0=2\pi/T_0$) such that $\tau_a/n_a+\tau_b/n_b=T_0/2$; this latter corresponds to a half-wave condition for  the unit cell of type ``$ab$''. Moreover, as is common in  photonic quasicrystal studies, we define the {\em structure parameter} (analogous to a duty cycle)
\beq
\gamma=\frac{\varphi_a}{\varphi_a+\varphi_b}.
\label{eq:gamma}
\eeq

%
\begin{figure}
	\centering
	\includegraphics[width=\linewidth]{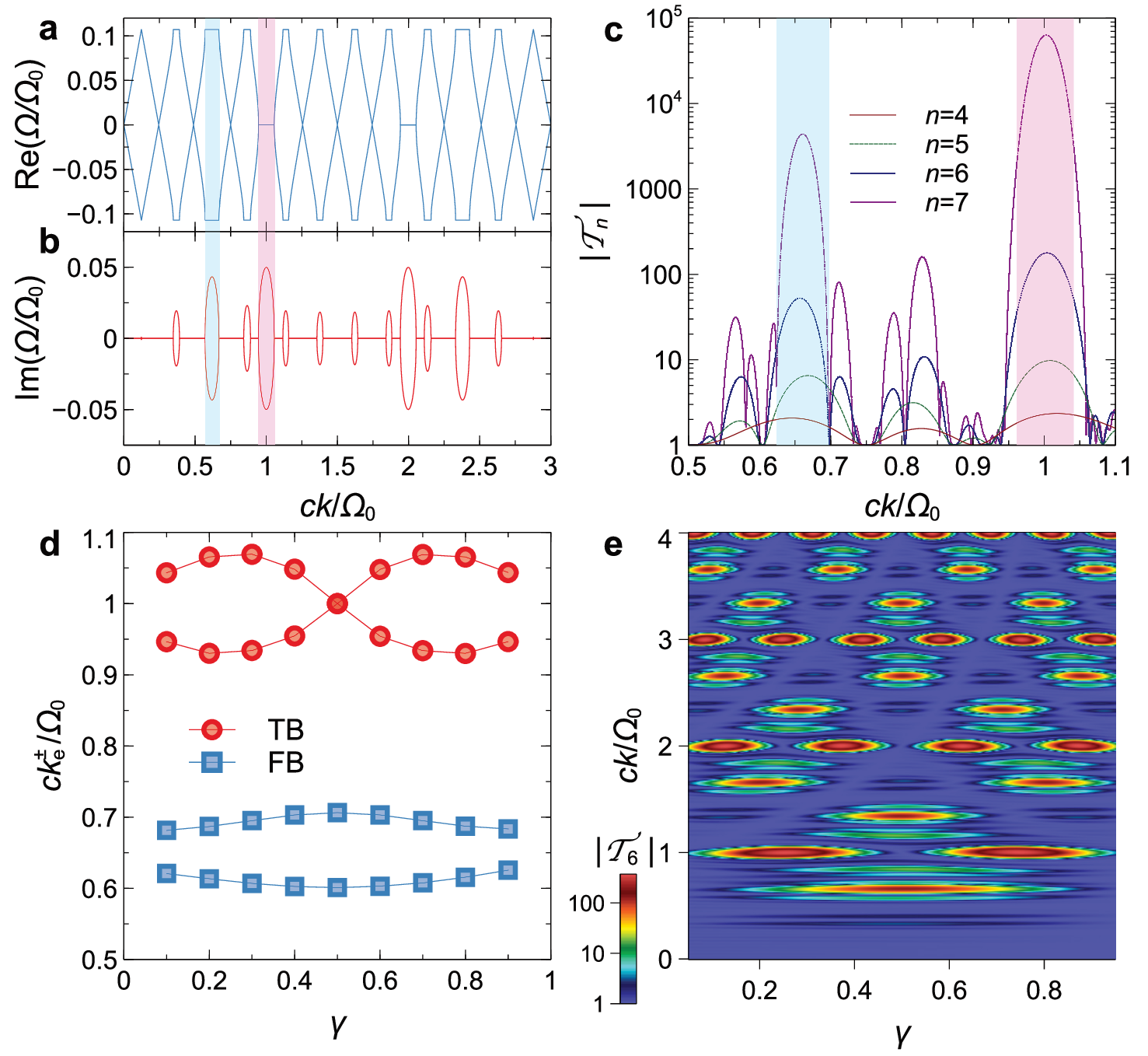}
	\caption{(a), (b) Band structure (real and imaginary part, respectively) for a Thue-Morse PTQC supercell with parameters $n_a=1$, $n_b=1.5$ and $\gamma=2/3$, at iteration order $n=4$. Due to symmetry and periodicity, only the interval $0\le ck/\Omega_0\le3$ is shown. The magenta- and blue-shaded regions identify a representative TG and FG, respectively. 
		(c) Transmission-coefficient magnitude nearby the representative TG (centered at $ck/\Omega_0=1$) and FG (centered at $ck/\Omega_0=2/3$), for various iteration orders; note the semi-log scale. (d) Band-edge normalized wavenumbers $k_e^{\pm}$ for the TG and FG, at iteration order $n=6$. (e) Transmission-coefficient magnitude in false-color log-scale as a function of normalized wavenumber and structure parameter, at iteration order $n=6$.}
	\label{Figure2}
\end{figure}

Figure \ref{Figure2} illustrates some representative results of our analysis. In particular, Figs. \ref{Figure2}a,b show the band structure corresponding to a supercell at iteration order $n = 4$ (i.e., $L_4 = 16$ modulation intervals) for a structure parameter $\gamma=2/3$. A complex pattern of momentum bandgaps emerges, with some gaps centered at integer values of the normalized wavenumber $ck/\Omega_0 = \left(\varphi_a + \varphi_b\right)/\pi$, reminiscent of those observed in conventional (periodic) PTCs. Adopting the terminology commonly used in the photonic quasicrystal literature \cite{Jiang:2005pg}, we identify these as ``traditional gaps'' (TGs). 

In addition to TGs, other momentum bangaps appear, centered at noninteger values of $ck/\Omega_0$ (e.g., 1/3, 2/3). 
These gaps, often termed ``fractal gaps'' (FGs) \cite{Jiang:2005pg}, are a defining characteristic of Thue-Morse PTQCs and, more broadly, a fundamental feature of aperiodic order.
To illustrate the difference, Fig. \ref{Figure2}c shows the evolution of the transmission coefficient of finite PTQCs at different iteration orders, with focus on a representative TG and FG, centered at $ck/\Omega_0=1$ and $2/3$, respectively. 
A markedly different behavior emerges: the TG exhibits a peak of increasing amplitude with iteration order, resembling the behavior observed in periodic PTCs, while the FG additionally develops increasingly complex patterns characterized by the appearance of multiple sidebands with {\em self-similar} nature (see \cite{Liu:1997po,Grigoriev:2010bm} for more details).
The dependence on the structure parameter also differs significantly between TGs and FGs, as illustrated in Fig. \ref{Figure2}d for the band edges $k_e^{\pm}$, defined as the upper and lower wavenumbers at which the transmission coefficient drops to unity) at iteration order $ n=6$. Specifically, the FG reaches its maximum width at $\gamma = 1/2$ (i.e., quarter-wave modulation intervals), whereas the TG is widest at $\gamma = 1/4$ and $\gamma = 3/4$, and vanishes entirely at $\gamma = 1/2$.
For a more comprehensive view, Fig. \ref{Figure2}e shows the transmission response as a function of both the normalized wavenumber and the structure parameter, for a fixed iteration order of $n = 6$. Additional results for various iteration orders are provided in \cite{SM}, where they are compared with the corresponding reference periodic PTQC case.

From this comparison, we observe a significantly denser spectrum in the PTQC case, with multiple tailoring mechanisms influenced by both the iteration order and the structure parameter. Notably, for rational values of the structure parameter $\gamma = j/l$, where $j, l \in \mathbb{N}^+$, $j < l$, and $j$ and $l$ are coprime, the spectrum exhibits periodicity with a period of $l$ in normalized wavenumber units; additionally, within each period, the spectrum is mirror-symmetric around its center. As the iteration order increases, the spectrum becomes increasingly complex, displaying self-similarity and a characteristic trifurcation \cite{Liu:1997po,Grigoriev:2010bm}. Asymptotically, it approaches a Cantor-like fractal structure, with gaps emerging at all scales \cite{Lei:2007pb} (see \cite{SM} for representative examples).

From a physical perspective, TGs arise from scattering at periodically correlated temporal boundaries within the Thue-Morse structure, akin to the bandgaps observed in periodic PTCs \cite{Lei:2007pb}. This correlation is uniformly distributed, meaning that the gap properties are independent of the sequence length. Mathematically, this implies that the TG edges are entirely determined by the second-order trace condition $\chi_2 = 2$ \cite{Lei:2007pb}. It can be shown (see \cite{SM} for details) that at the center of the $m$th TG, the second-order trace is given by:
\beq
\chi_2=
\left[1
-\frac{1}{2}\left(
\frac{n_a}{n_b}+\frac{n_b}{n_a}
\right)
\right]\cos\left(
4m\pi\gamma
\right)+1
+\frac{1}{2}\left(
\frac{n_a}{n_b}+\frac{n_b}{n_a}
\right).
\label{eq:chi2gamma}
\eeq
From this expression, it is clear that if $\gamma = j/(2m)$, with $j \in \mathbb{N}^+$, then $\chi_2 = 2$, causing the TG to vanish. Therefore, for example, when $\gamma = 1/2$, the spectrum is entirely devoid of TGs.

%
\begin{figure}
	\centering
	\includegraphics[width=.8\linewidth]{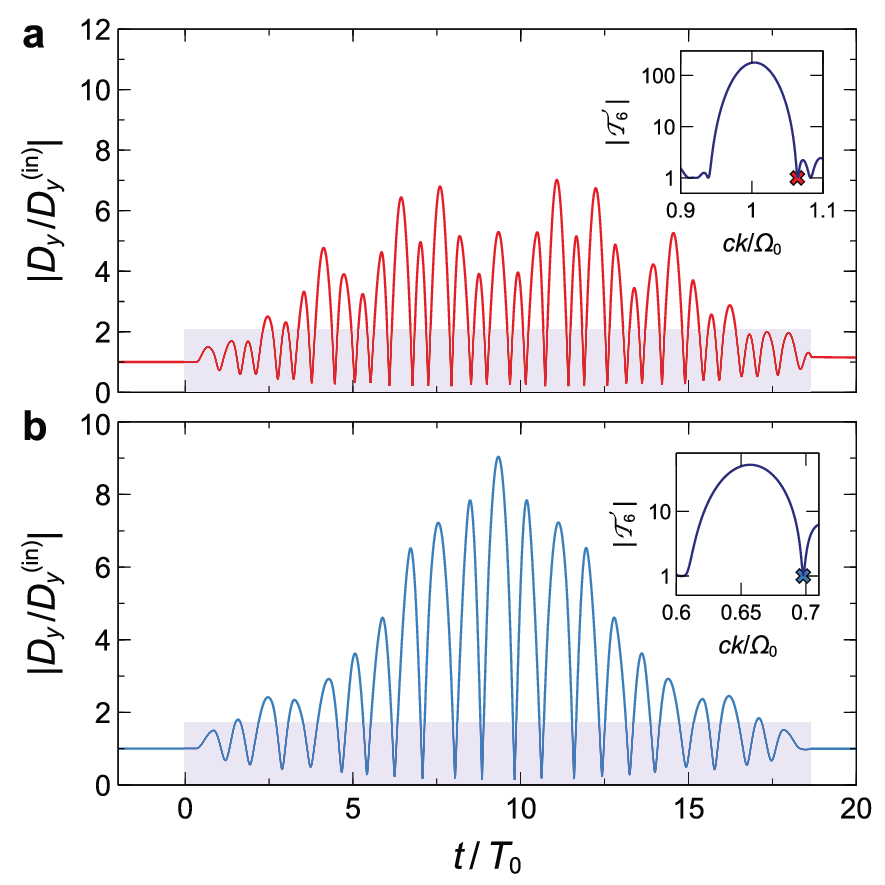}
	\caption{Examples of gap-edge states. (a) Electric displacement field magnitude (normalized with respect to incident one) as a function of time, for a Thue-Morse PTQC with parameters $n_a=1$, $n_b=1.5$ and $\gamma=2/3$, at iteration $n=6$ and normalized wavenumber $ck/\Omega_0=1.064$ (corresponding to a TG edge). (b) Same as in panel (a), but at $ck/\Omega_0=0.698$ (corresponding to a FG edge). The purple shaded bars identify the interval where  temporal modulation is active. The insets show the transmission responses, with the chosen wavenumbers identified by cross-shaped markers.}
	\label{Figure3}
\end{figure}

Conversely, FGs emerge from the multi-scale nested structure of temporal boundaries within the Thue-Morse sequence. The underlying correlations are more intricate and length-dependent compared to those for TGs, involving three distinct partitioning schemes, with one scheme being dominant \cite{Lei:2007pb}. More recently, their origin has also been linked to the evolution of certain topological singularities \cite{Zhang:2021ft}.

The fundamental difference between TGs and FGs is also evident in the nature of their gap-edge states. In TGs, these states are {\em extended}, exhibiting Bloch-like characteristics similar to those found in periodic PTCs, as shown in Fig. \ref{Figure3}a. Conversely, in FGs, the gap-edge states are {\em quasi-localized}, as illustrated in Fig. \ref{Figure3}b. While state localization can also occur at the temporal boundary between different periodic PTCs \cite{Lustig:2018ta}, as well as in the presence of random temporal disorder \cite{Eswaran:2024al}, the nature of FG edge states is fundamentally distinct, as they originate from the deterministic long-range order inherent to the system, and do not generally exhibit exponential localization. As demonstrated in \cite{SM}, these states acquire a fractal structure with increasing iteration order, characterized by large magnitude fluctuations and cluster-periodic patterns. This distinctive behavior enables anomalous localization effects and advanced pulse-shaping capabilities that are unattainable in periodic or random PTCs.

%
\begin{figure}
	\centering
	\includegraphics[width=.8\linewidth]{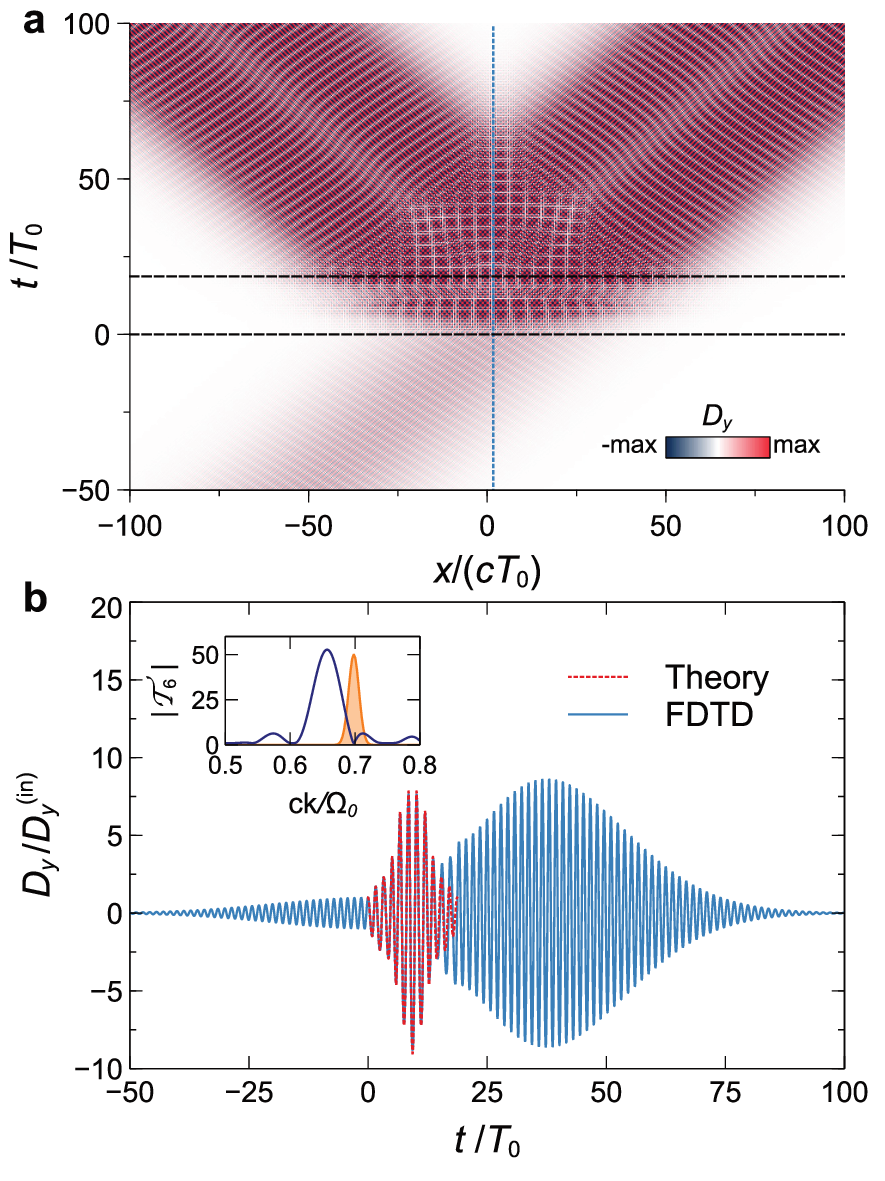}
	\caption{(a) FDTD space-time map of  electric displacement field for a Thue-Morse PTQC with parameters $n_a = 1$, $n_b = 1.5$, and $\gamma = 2/3$, at iteration order $n = 6$. The system is excited by a narrowband Gaussian wavepacket with a spectrum centered at $ck/\Omega_0 = 0.698$ [see inset in panel (b)]. The horizontal black-dashed lines indicate the modulation interval.  	
		(b) Temporal field profile at $x = 1.70 cT_0$ [marked by the vertical blue-dotted line in panel (a)], comparing the FDTD simulation results with theoretical predictions (real part of the wavefield from Fig. \ref{Figure3}b within the modulation interval). The inset displays the transmission response, with the Gaussian wavepacket spectrum overlaid in orange shading.}
	\label{Figure4}
\end{figure}

For independent verification of our theoretical predictions, Fig. \ref{Figure4} presents results from finite-difference time-domain (FDTD) full-wave simulations \cite{Oskooi:2010ma} (see also \cite{SM} for details). In this example, the system is excited by a narrowband Gaussian wavepacket centered at the FG edge, $ck/\Omega_0 = 0.698$ (see inset in Fig. \ref{Figure4}b).  
Figure \ref{Figure4}a displays the space-time map of the electric displacement field, revealing two key phenomena: {\em (i)} field localization within the temporal modulation region, and ({\em ii}) pulse splitting accompanied by amplification. These effects are further quantified in Fig. \ref{Figure4}b, which shows a temporal cut at a fixed position in space. As the temporal modulation begins, the FG edge state clearly emerges, demonstrating excellent agreement with our theoretical, time-harmonic prediction (shown only within the modulation interval). Subsequently, amplification occurs due to spectral components of the wavepacket falling within the gaps. 
As shown in Fig. \ref{Figure4}b, the field continues to grow for a short time even after the temporal modulation is switched off. This behavior can be explained by the space-time map in Fig. \ref{Figure4}a, which reveals an interference effect between forward and backward propagating waves near the origin ($x=0$), resulting in a localized enhancement of the field amplitude.

As shown in \cite{SM}, when the wavepacket spectrum lies predominantly within a gap, the response closely resembles that observed in periodic PTCs.


In conclusion, we have developed a theoretical framework for analyzing PTQCs, using the Thue-Morse sequence as a representative model of aperiodic temporal order. By extending the trace and anti-trace map formalism to the temporal domain, we have systematically explored their band structure and wave-transport properties. This approach provides valuable insights into the underlying physics, spectral characteristics, and scaling behavior of PTQCs, effectively parameterizing the influence of key design variables (structure factor and iteration order) on the spectral response across multiple scales.

Our findings suggest that aperiodic temporal order can serve as a useful and flexible degree of freedom, offering new possibilities for spectral control and localization in photonic systems.
The broadband and complex spectral response of PTQCs makes them promising candidates for applications such as lasing and pulse shaping, offering greater flexibility in phase-matching and waveform design.  

A crucial design consideration in this framework is the modulation sequence. While our analysis has focused on a specific example, many of the observed effects are broadly applicable. For instance, \cite{SM} presents a similar study on PTQCs based on Fibonacci sequences \cite{Queffelec:2010sd}, another widely studied model in spatial quasicrystals \cite{Kohmoto:1987lo, Gellerman:1994lo, Macia:1998oe, DalNegro:2003lt, Jagannathan:2021tf}. These additional results further reinforce the key trends and general principles highlighted in our work.

Importantly, our framework is adaptable to {\em arbitrary} sequences, enabling the systematic exploration of the vast intermediate regime between perfect periodicity and randomness, including deterministic disorder. This flexibility not only opens new possibilities for targeted bandgap engineering in photonic applications, but also provides a pathway  to a deeper understanding of the complex interplay between order, disorder, and wave dynamics in time-modulated systems.

Finally, although our findings are derived in a photonic context, their implications are much broader, as time-modulation of constitutive parameters is gaining importance across various physical fields, including acoustic \cite{Chong:2024mi}, elastic \cite{Trainiti:2019tp}, and thermal \cite{Li:2022ro} metamaterials.

~\\
\begin{acknowledgments}
{\em Acknowledgments} --
	This work was supported in part by the European Union-Next Generation EU under the Italian National Recovery and Resilience Plan (NRRP), Mission 4, Component 2, Investment 1.3, CUP E63C22002040007, partnership on ``Telecommunications of the Future'' under Grant PE00000001 - program ``RESTART'', and in part from the University of Sannio via the FRA 2024 program.
\end{acknowledgments}


%

\end{document}